# Structural controllability of complex networks based on preferential matching

Xizhe Zhang [1*], Tianyang Lv [2,3,4], XueYing Yang[1], Bin Zhang[1]

[1] （College of Information Science and Engineering, Northeastern University, Shenyang110819, China）

[2] （College of Computer Science and Technology, Harbin Engineering University, Harbin 150001, China）

[3] （College of Computer Science and Technology, Tsinghua University, Beijing 100084, China)

[4] （Audit Research Institute, National Audit Office, Beijing 100830, China)

* Email: zhangxizhe@ise.neu.edu.cn

**Abstract**

Minimum driver node sets (MDSs) play an important role in studying the structural controllability of complex networks. Recent research has shown that MDSs tend to avoid high-degree nodes. However, this observation is based on the analysis of a small number of MDSs, because enumerating all of the MDSs of a network is a #P problem. Therefore, past research has not been sufficient to arrive at a convincing conclusion. In this paper, first, we propose a preferential matching algorithm to find MDSs that have a specific degree property. Then, we show that the MDSs obtained by preferential matching can be composed of high- and medium-degree nodes. Moreover, the experimental results also show that the average degree of the MDSs of some networks tends to be greater than that of the overall network, even when the MDSs are obtained using previous research method. Further analysis shows that whether the driver nodes tend to be high-degree nodes or not is closely related to the edge direction of the network.

**Introduction**

Controlling complex systems is a critical topic in many applications. A system is called controllable if it can be driven from any initial state to any desired state in a finite time. Previous researches have usually adopted a complex network as the fundamental model to analyze the topological structure[1-3], the evolving model[4-6], and the dynamic behavior[7-9] of complex systems.

However, we still lack a thorough understanding of how to control complex networks. According to the control theory, a linear time-invariant system whose states are determined by the following equation:

$$\frac{dx(t)}{dt} = Ax(t) + Bu(t) \tag{1}$$

where the vector $x(t)=(x_1(t), …, x_N(t))^T$, denotes the state of $N$ nodes in the network at time $t$, $A$ is the transpose of the adjacency matrix of the network, $B$ is the input matrix that defines how control signals are inputted to the network, and $u(t)=(u_1(t), …, u_H(t))^T$ represents the $H$ input signals at time $t$. A node whose control signal is directly inputted is called a driver node. The minimum sets of driver nodes to control a network are called the minimum driver nodes sets (MDSs).

Lin[10] presented a network representation of linear time-invariant systems and stated that the system is structurally controllable if and only if the network can be spanned by cacti structures. Commault[11] proved that the minimal signals need to control a network can be obtained by maximal matching[12] of network. Based on above works, Liu[13] developed an analysis tool to study the controllability of an arbitrary complex directed network, and found that MDSs tend to be composed of low-degree nodes in both real and model networks.

However, the maximum matching of a network is usually not unique[14], and thus neither are the MDSs. Previous studies[15-19] have only randomly sampled MDSs and analyzed a small number of the MDSs of a network because enumerating all possible maximum matchings is in the class of #P problem[20]. Therefore, the past researches have not been sufficient to arrive at a convincing conclusion about whether MDSs tend to avoid high degree nodes or not.

In this paper, we propose a preferential matching algorithm to find some MDSs with desired degree properties. To find these MDSs, the algorithm arranges the matching order of the nodes according to their degree rank. Because low-ranking nodes have higher probabilities of being driver nodes, the obtained MDSs tend to be composed of the high- or the medium-degree nodes of the network. The algorithm can also be applied to obtain the MDSs with other topological properties.

By using the preferential matching algorithm, we found that there were some MDSs composed mainly of high- and medium-degree nodes in some networks. Moreover, in some networks, the average degree of the MDSs tended to be greater than that of the overall network, even if the MDSs were obtained using the previous random-matching method.

We conclude that there are networks that favor low-degree MDSs and other networks that favor high-degree MDSs. To find the underlying reason for this phenomenon, we designed a directed BA model for model networks and a reversal strategy for the edge direction for real networks. The experimental results showed that the MDSs of the network tended to be composed of high-degree nodes if the majority of the edges of a network were pointing from high-degree nodes to low-degree nodes; otherwise, the MDSs of the network tended to be composed of low-degree nodes. Therefore, whether the driver nodes tended to be high degree or not was closely related to the edge direction of the network.

**Preferential Matching Algorithm**

First, we will briefly introduce the basic concepts of maximum matching. For a directed network $G$, $V(G)$ is the node set and $E(G)$ is the edge set, with $N=|V|$ and $L=|E|$. A set of edges in $G$ is called a matching $M$ if no two edges in $M$ have a node in common. A node $v_i$ is matched by $M$ if there is an edge of $M$ pointing to $v_i$, otherwise $v_i$ is unmatched. A path $P$ is said to be $M$-alternating if the edges of $P$ are alternately in and not in $M$. An $M$-alternating path $P$ that starts and ends at the unmatched nodes is called an $M$-augmenting path. A matching with the maximum number of nodes is called a maximum matching $M^*$. A matching $M$ is called a perfect matching if all of the nodes of $G$ are matched by $M$.

The minimum input theorem[38] proves that if there is a perfect matching in a network, the number of driver nodes is one, otherwise the number of driver nodes is equal to the number of unmatched nodes with respect to any maximum matchings. And the driver nodes are unmatched nodes. The size of the maximum matching $M^*$ is denoted $|M^*|$. The minimum number of driver nodes is thus

$$n_D = \max\left\{N - |M^*|, 1\right\} \qquad (2)$$

Based on this theorem, the MDSs can be obtained by finding the maximum matchings of a network. Therefore, it is critical to find all of the maximum matchings. Previous maximum matching algorithms, such as Hopcroft-Karp[12] and the Hungarian algorithm[21], are based on the theorem proposed by Berge[22]. That theorem proves that $M^*$ is a maximum matching if and only if there is no augmenting path in $G$ relative to $M^*$. Therefore, the basic idea of the maximum matching algorithm is as follows: first, find an augmenting path from each unmatched node by current matching $M$ (initially $M=\varphi$), then obtain an expanded matching $M'$. Repeat the first and the second steps until no augmenting path exists. The final matching is a maximum matching. Using this process, once a node $v_i$ becomes a matched node, it will be matched by the final maximum matching and won't be a driver node.

Therefore, if we deliberately arrange the matching order of nodes according to the order of degree, we would find MDSs with a desired degree property such as finding some high-degree MDSs, particularly when a network has many maximum matchings. However, the matching order of nodes is determined by the time when a node first appears in the augmenting path, but the time is hard to be pre-decided. It is possible that a node with a high degree appears very early in an augmenting path, even if it is arrange to be the last one as the start of augmenting paths. For example, we can sort the nodes as $\{v_0,v_1,v_2,v_3,v_4,v_5,v_6\}$ in the ascending order by degree and treat this order as the input sequence to select the unmatched start node in finding an augmenting path. But we may find an augmenting path $P$ $v_0 \rightarrow v_4 \rightarrow v_5 \rightarrow v_6$ at the very first step. Although the path starts from $v_0$ with the lowest degree, it contains the highest degree nodes $v_4$, $v_5$ and $v_6$ and these nodes cannot be the driver nodes of the final MDSs. Thus, the matching order of the nodes would be quite different from the degree order of the nodes, and the MDSs with a desired degree property could not be easily found.

To overcome this problem, we designed an iterative preferential matching method. We sort the nodes as $\{v_0, v_1,...v_n\}$ in the ascending order by degree and denote $m$ as the number of preferential matching nodes. The method starts from the sub graph $H_0$ with the lowest-degree node ranked first; at each iterative step $i$, the sub graph $H_i$ will be extended by adding the node with the $i$-th rank, and the maximum matching of $H_i$ is calculated based on the previously obtained maximum matching of $H_{i-1}$. We repeat this procedure until the sub graph $H_i$ is equal to the whole network or until $m$ preferential nodes have been added. Details of the preferential matching method are as follows:

1. Sort nodes as $\{v_0, v_1,...v_n\}$, $H_0=\{v_0\}$, $M^*_0=\varphi$, $i=1$;
2. Set $H_i = H_{i-1} + \{v_i\}$ and find a maximum matching $M^*_i$ of $H_i$ based on $M^*_{i-1}$, $i=i+1$;
3. Repeat step 2 until $i=m$;
4. If $m<N$, find the resulting maximum matching $M^*$ of $G$ based on $M^*_m$; else $M^*_m$ is the resulting maximum matching of $G$, and the MDS is composed of the unmatched nodes with respect to $M^*$.

An example of the proposed method is shown in Figure 1.

We obtain a maximum matching of $G$ in the step 4. And, as with current algorithms[12, 21], once $v_i$ is matched in the process, it must be matched by the resulting maximum matching. The proposed method ensures that we can find the maximum number of matched nodes of $H_i$ from the first $i$ ranking nodes and that a high-degree node will not be matched in early steps because the node is not included in the early sub-graphs. Therefore, we can make the matching order of the nodes as similar as possible to the predefined order of degrees. Thus, high-degree nodes will have a higher probability of being the driver nodes. However, the order of arrangement has no influence on some particular nodes, for instance the nodes with zero in-degree must be driver nodes no matter what the input order is.

**Experimental Results and Analysis**

To analyze the degree property of MDSs, we selected 21 real networks that belong to 12 categories, including trust networks, food networks, electric networks, neuronal networks, citation networks, the World Wide Web, the internet, social communication networks and social organization networks. Table 1 shows the average degree of a network $<k>$, the size of the networks' MDSs $n_D$, and the fraction of driver nodes $\lambda_D = n_D/N$.

First, we find the MDSs with the desired high-degree property based on the preferential matching algorithm. Let $<k^D>$ be the average degree of the MDSs obtained under a different number $m$ of preferential nodes, and let $<k^D_{max}>$ and $<k^D_{min}>$ be the maximum and the minimum $<k^D>$ of all of the obtained MDSs, respectively. Figure 2 shows the variation in $<k^D>$ versus $m$ in the real and model networks. Obviously, the preferential matching method can find MDSs with the preferred high-degree property, and the high-degree property becomes clearer with the increment of $m$. If the nodes are sorted in ascending order according to degree, $<k^D>$ will increase with $m$ to the upper bound $<k^D_{max}>$; if the nodes are sorted in descending order according to degree, $<k^D>$ will decrease with $m$ to the lower bound $<k^D_{min}>$.

From Table 1 and Figure 2, a basic observation was that the MDSs were structurally diverse: the $<k^D>$ of many networks varied widely. Thus, the different MDSs of the same network could have quite different degree properties. Moreover, $<k^D_{max}>$ was greater than $<k>$ in many networks, such as the *Grassland*, *Seagrass*, *Ythan*, and *Florida* networks. Therefore, we were able to find the MDSs whose $<k^D>$ was greater than the average degree of the network.

To further verify the above observation, we analyzed the degree distribution of driver nodes of the MDSs with high $<k^D>$. We computed the MDS with the highest average degree $<k^D_{max}>$ by using the preferential matching method. Figure 3 shows the results of some real and model networks. In Figure 3, each point corresponds to the set of nodes with the specific degree $k$. The black point means that no node with the degree $k$ appears in the result MDS, and the red point means that some nodes with the degree $k$ appear in the result MDS. The inset graph shows the degree distribution of all driver nodes of the MDS with $<k^D_{max}>$. Therefore, if all red points have high degree, the MDS tends to be composed of high-degree nodes. It can be seen from Figure 3 that there do exist the MDS mainly composed of high- or medium- degree nodes in some networks. Taking the *world-trade*[38] network as an example, 66.2% of its nodes have $k \leq 20$, but none of these low-degree nodes appeared in the result MDS; meanwhile, 88.9% of the rest high-degree nodes with $k > 20$ appeared in the MDS. Similar results can be observed in the *BA* and *ER* networks. However, not all networks had the MDS mainly composed of high-degree nodes. The MDS with $<k^D_{max}>$ of some networks was composed of the nodes with degree ranging from the lowest degree to the highest, such as the *seagrass*[26], *florida*[27] and *c.elegans*[29] networks, while the MDS with $<k^D_{max}>$ of other networks was mainly composed of the low-degree nodes, such as the *P2P-1*[33] network.

Second, we tried to prove that the average degree of the MDSs of some networks tended to be greater than that of the overall networks, even if the MDSs were obtained using the previous random matching method. In the experiment, we randomly sampled 10,000 different MDSs of each network. Table 1 shows the average value $\overline{\langle k^D \rangle}$ of the average degree of all of the sampled MDSs because the average degree of the different MDSs varied. We found that the $\overline{\langle k^D \rangle}$ of some networks, such as the *Zewail*, *world trade* and *literature* networks, were greater than or equal to $<k>$ even when using the previous sample method[13].

Finally, these experimental results provoked us to explain why the driver nodes of some networks tended to be low degree while others were not. According to the minimum input theorem, a driver node is not pointed to by any matched edge. Therefore, if the majority of edges of a network point from high-degree nodes to low-degree nodes, the MDSs tend to be composed of high-degree nodes. Otherwise, the MDSs tend to be composed of low-degree nodes. Figure 4 gives an example where two networks have the same topology except that the directions of their edges are opposite. The edges of the network in Figure 4(*a*) are pointing to the low-degree nodes, while the edges in Figure 4(*b*) are pointing to the high-degree nodes. Therefore, they have very different MDSs. The driver nodes of network Figure 4(*a*) are $v_1$, $v_3$ and $v_4$ and have the highest degrees, while the driver nodes of network Figure 4(*b*) are $v_5$, $v_6$ and $v_7$, which have the lowest degrees.

Therefore, we believe that the node composition of the MDSs is closely related to the direction of the edges in a network. To verify this hypothesis, we designed a revised BA model to generate directed networks. The model was the same as the classical BA model [39] except that the direction of a newly added edge is determined by the following rule: the direction of the new edge points from an existing old node $v_{old}$ to a new node $v_{new}$ with probability $p$, and the probability of pointing in the opposite direction is 1-$p$. Therefore, if $p$ is large enough, the edges of a high-degree node $v_{old}$ will have a high probability of pointing to other nodes. The result of this arrangement is that the edges of a generated network tend to point from high-degree nodes to low-degree nodes, so the high degree nodes are more likely to be the source nodes[42], which must receive the control signal from outside. We calculated the fraction $f_{hi-lo}$ of edges that pointed from high-degree nodes to low-degree nodes in a directed BA network. Figure 5(a) shows the linear relation between $f_{hi-lo}$ and $p$.

Then, we randomly calculated 10,000 MDSs of several directed BA networks using the Hopcroft-Karp algorithm. Figure 5(b) shows the average degree of the MDSs $\overline{\langle k^D \rangle}$ increases with $p$. When $p$=0.5, which means that the direction of the edges are randomly decided, $\overline{\langle k^D \rangle}$ is much less than <$k$>; as $p$ increases to close to 1, $\overline{\langle k^D \rangle}$ gradually becomes greater than <$k$>; and in Figure 5(c), when $p$=1, the $\overline{\langle k^D \rangle}$ of all of the directed BA networks is always greater than <$k$>.

We also verified this hypothesis in the real networks. Due to the complexity of degree correlation in real directed networks[40], there may be no obvious relationship between $\overline{\langle k^D \rangle}$ and $f_{hi-lo}$ in different real networks. Therefore, we designed the following edge-reversal strategy to verify this hypothesis: for an edge $v_i \rightarrow v_j$, if $k_i < k_j$, then reverse the edge direction to $v_j \rightarrow v_i$ with probability $R$. Similarly to the directed BA model, if $R$ is large enough, the edges of a high-degree node will have a high probability of pointing to a low-degree node. Figure 5(d) shows $\overline{\langle k^D \rangle}$ versus $R$. We can see that if the original $\overline{\langle k^D \rangle}$ of a network is less than <$k$>, the $\overline{\langle k^D \rangle}$ increases gradually with the increase of $R$ and becomes greater than or equal to the <$k$> of the network. However, for a few networks such as *TRN-Yeast*-1, the average degree of the MDSs will decrease with $R$. This finding suggests that other topological factors also influence the degree properties of MDSs, although the direction of the edges may be a major factor.

**Discussion**

The minimal driver nodes set can be obtained by finding the maximal matching of network. However, the MDSs of a network are not unique, and have very different topological features exist. Thus, one important research direction in the controllability of complex networks is analyzing the topological features of all of the possible MDSs.

However, enumerating all of the MDSs is in the class of #P problem, so we tried to find the MDSs with specific topological features. Our contribution in this paper was twofold. First, we proposed a MDS-discovery method based on preferential matching. This method could effectively find a MDS with a high average degree by arranging the matching sequence of nodes based on the order of their degree. Furthermore, we were able sort nodes by any desired property and found a MDS satisfying that property. The algorithm also showed the promise for finding a MDSs that satisfy application-specific constraints. For instance, if some nodes cannot be driver nodes in practice, we let these nodes be matched with high priority in the preferential matching process; thus, a MDS without these nodes can be obtained if such a MDS exists.

Second, we found that whether driver nodes tended to be low degree was closely related to the direction of edges. If the majority of edges pointed to low-degree nodes, control signals were required to transfer from high-degree nodes to low-degree nodes; thus, the MDSs tended to be composed of high-degree nodes.

Future research will investigate all of the possible MDSs and analyze the degree distribution of the driver nodes of networks. In this manner, we may discover an optimal strategy for finding MDSs that satisfy specific constraints.

Table 1 | Overview of real networks and the statistical results of their MDSs. $<k>$ is the average degree of a network, $n_D$ is the size of a MDS, $\lambda_D = n_D/N$, $<k^D>$ is the average degree of the MDS, $\overline{\langle k^D \rangle}$ is the average value of $<k^D>$ for all of the obtained MDSs, and $<k^D_{min}>$ and $<k^D_{max}>$ are the maximum and the minimum values $<k^D>$ of all of the MDSs obtained by the preferential matching method under a different preferential matching number $m$.

| Type | Name | $N$ | $L$ | $<k>$ | $\overline{\langle k^D \rangle}$ | $[<k^D_{min}>, <k^D_{max}>]$ | $n_D$ | $\lambda_D$ |
|---|---|---|---|---|---|---|---|---|
| Trust | Wiki-Vote[23] | 7115 | 103689 | 29.15 | 9.66 | [9.66,9.66] | 4736 | 0.67 |
| Food Web | Grassland[24] | 88 | 137 | 3.11 | 2.67 | [2.20,3.02] | 46 | 0.52 |
| | Little Rock[25] | 183 | 2494 | 27.26 | 15.39 | [14.79,15.83] | 99 | 0.54 |
| | Seagrass[26] | 49 | 226 | 9.22 | 8.06 | [6.46,11.08] | 13 | 0.27 |
| | Ythan[24] | 135 | 601 | 8.90 | 7.43 | [4.86,9.67] | 69 | 0.51 |
| | Florida[27] | 128 | 2106 | 32.91 | 24.86 | [16.1,36.6] | 30 | 0.23 |
| | Mondego[28] | 46 | 400 | 17.39 | 12.47 | [9.26,12.58] | 19 | 0.41 |
| Power Grid | USpowerGrid[29] | 4941 | 13188 | 10.68 | 2.73 | [2.06,3.50] | 575 | 0.12 |
| Neuronal | C. elegans[29] | 306 | 2345 | 15.33 | 5.6 | [3.36,16.47] | 58 | 0.19 |
| Citation | Hep-th[30] | 27770 | 352807 | 25.41 | 9.45 | [7.96,11.64] | 5994 | 0.22 |
| | Zewail[31] | 6752 | 54233 | 16.064 | 17.55 | [4.99,25.02] | 2427 | 0.36 |
| | Kohonen[31] | 4470 | 12731 | 5.696 | 5.73 | [2.67,6.3] | 2812 | 0.63 |
| WWW | Polblogs[32] | 1224 | 16718 | 27.32 | 12.41 | [4.12,17.41] | 418 | 0.34 |
| Internet | P2P-1[33] | 10876 | 39994 | 7.36 | 6.92 | [2.67,9.45] | 6004 | 0.55 |
| Social-Communication | UCIonline[34] | 1899 | 20296 | 21.38 | 6.75 | [1.72,13.76] | 614 | 0.32 |
| Regulatory | TRN-Yeast-1[35] | 4441 | 12873 | 5.80 | 5.85 | [3.18,5.95] | 4284 | 0.96 |
| Companies | Eva[36] | 8497 | 6726 | 1.584 | 1.59 | [1.36,1.61] | 7194 | 0.85 |
| Literary | literature[37] | 35 | 81 | 4.628 | 4.72 | [4.46,5.46] | 13 | 0.37 |
| Trade | World_trade[38] | 80 | 998 | 24.95 | 26.93 | [10.46,47.33] | 24 | 0.3 |

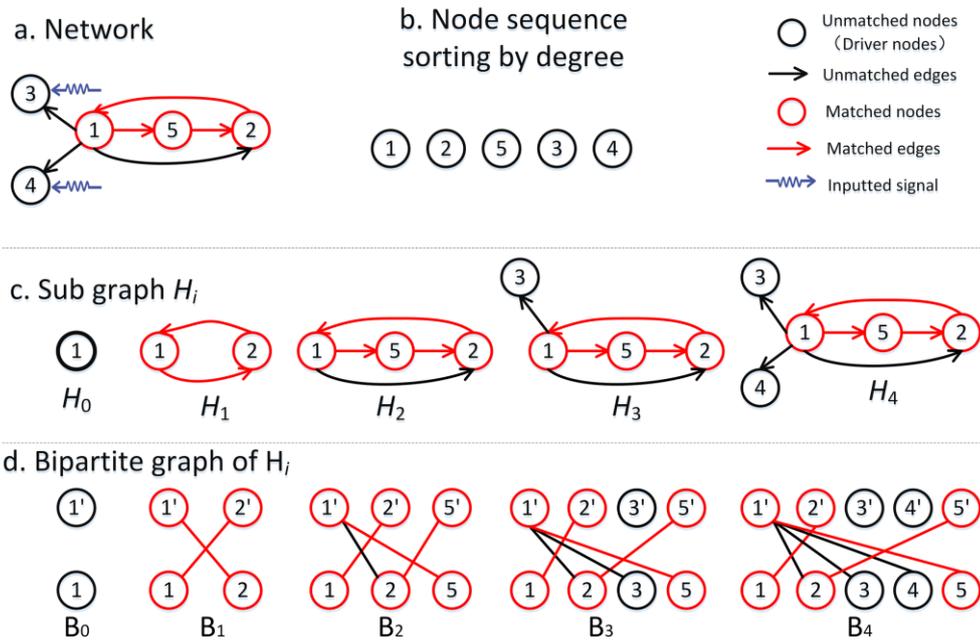

Figure 1 | Illustration of the preferential matching algorithm process. We rank all of the nodes in descending order by degree, and the driver nodes are nodes $v_3$ and $v_4$ that are the last two of the sequence.

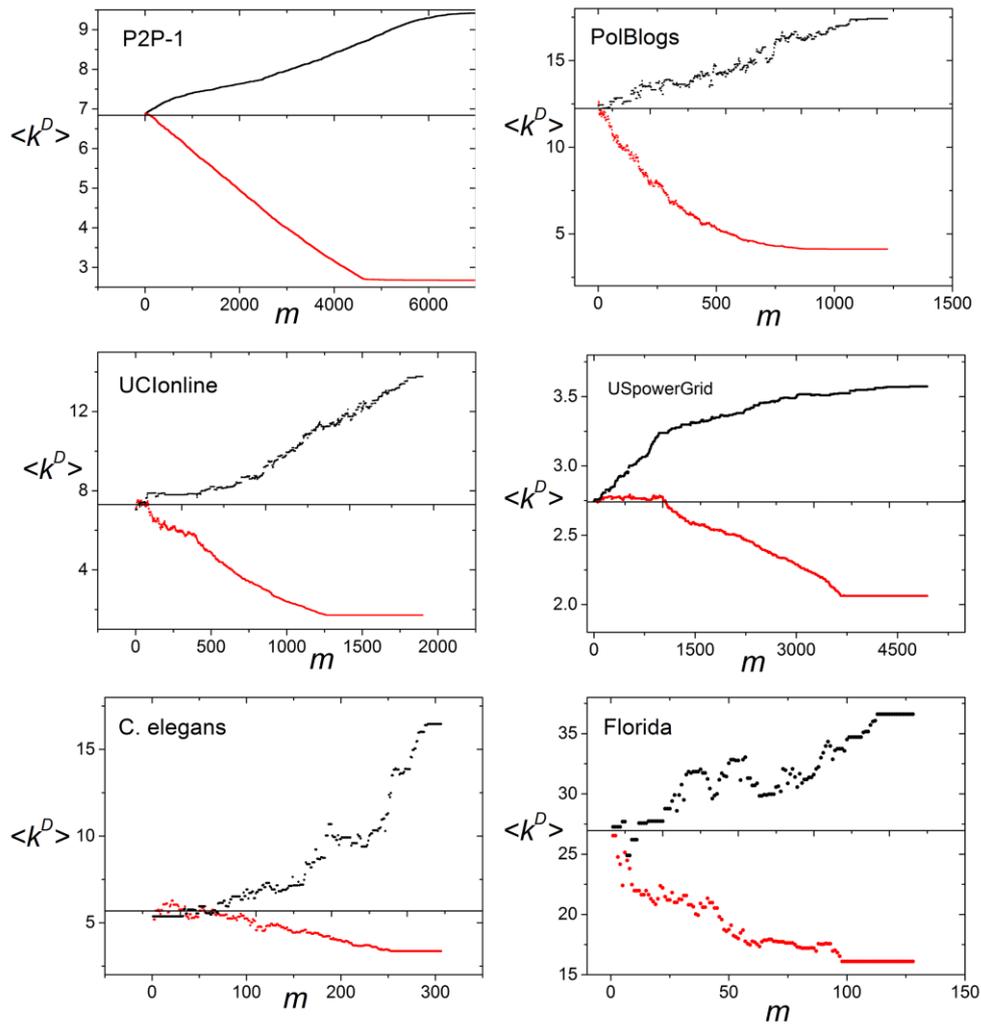

Figure 2 | Relationship of $<k^D>$ of a MDS versus the preferential matching number $m$. The results above the solid line show the value of $<k^D>$ when nodes are sorted in ascending order by degree. The results below the solid line show the value of $<k^D>$ when nodes are sorted in descending order by degree.

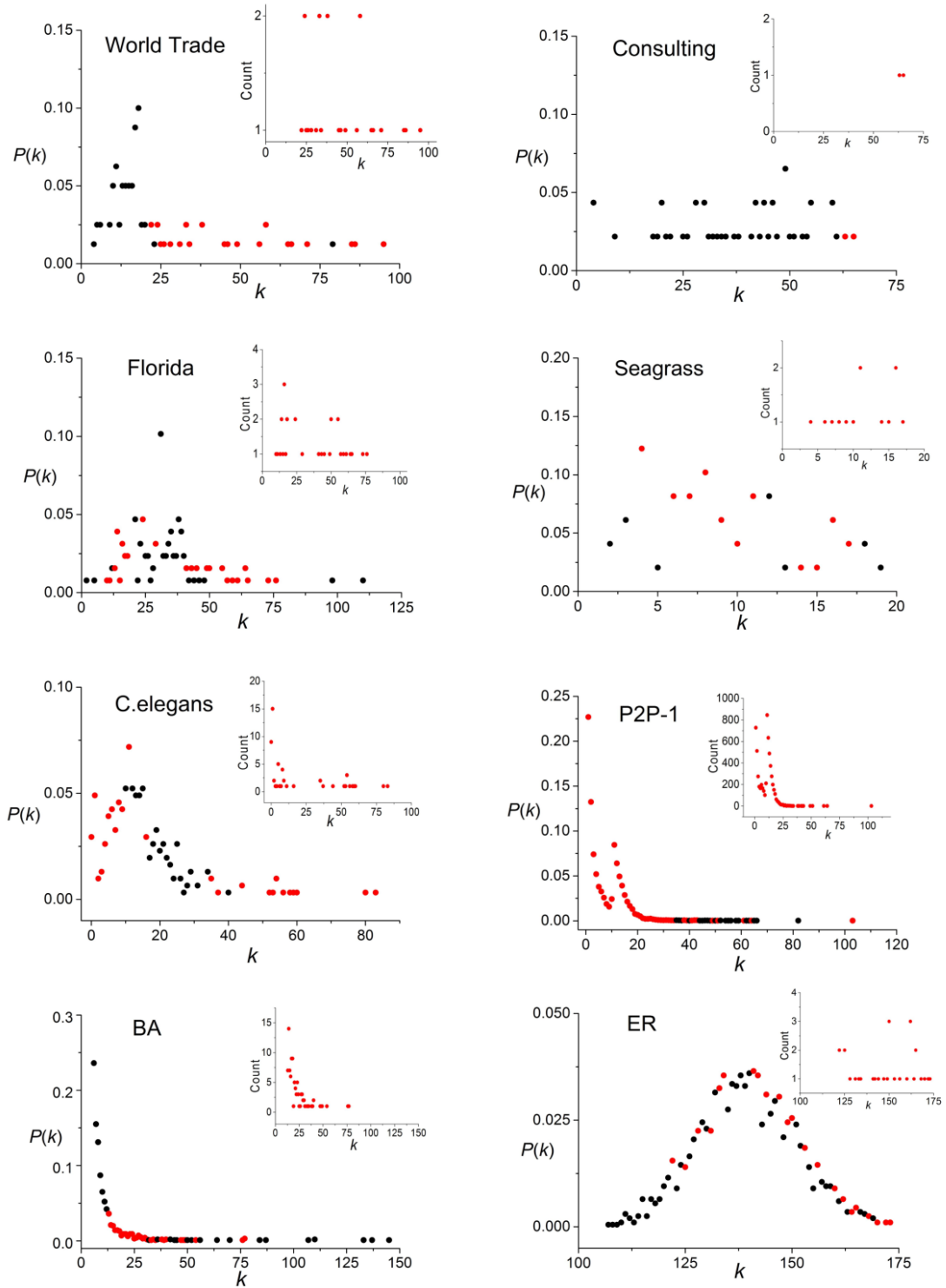

Figure 3 | Degree distribution of driver nodes in real and model networks. The MDS with the highest average degree $<k^D_{max}>$ was computed by using the preferential matching method. Each point corresponds to the set of nodes with the specific degree $k$, the black point means that no node with the degree $k$ appeared in the result MDS and the red point means that some nodes with the degree $k$ appeared in the result MDS. The inset graph shows the degree distribution of all driver nodes of the MDS with $<k^D_{max}>$.

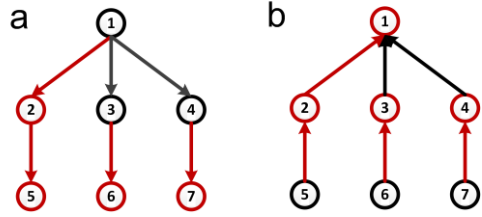

Figure 4 | Two simple networks with *<k>*=1.857. Red nodes and edges are matched by a maximum matching. Black nodes and edges are driver nodes and unmatched edges. The average degrees of the MDSs of networks (a) and (b) are 2.33 and 1, respectively.

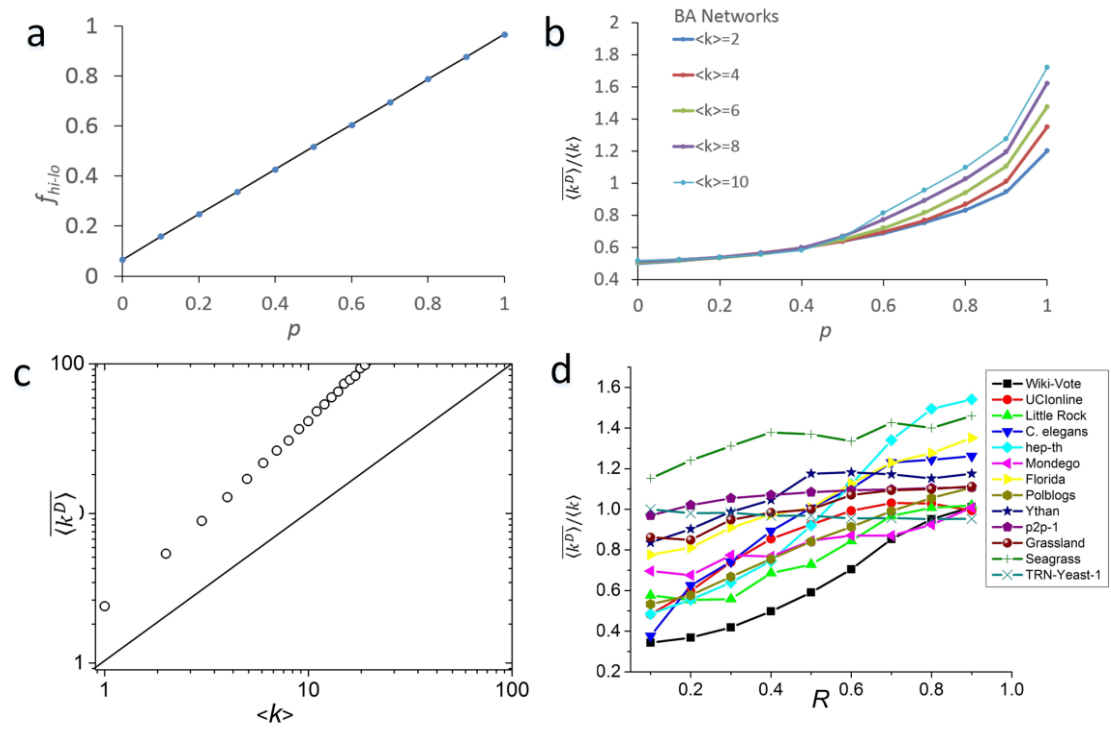

Figure 5 | Edge direction strongly influences the average degree of MDSs. (a) The fraction $f_{hi\text{-}lo}$ and the probability *p* have a clear linear relation in directed BA networks; (b) the ratio of $\overline{\langle k^D \rangle}$ and *<k>* increase with *p* in directed BA networks; (c) the $\overline{\langle k^D \rangle}$ of all directed BA networks is always greater than *<k>* when $p = 1$; (d) the ratio of $\overline{\langle k^D \rangle}$ to *<k>* increases with the reversal probability *R* in real networks.